# Graphical Query Builder in Opportunistic Sensor Networks to discover Sensor Information

A.F.M. Sultanul Kabir[1] and Mohammad Saiful Islam Mamun[2]

[1]Department of Computer Science, American International University Bangladesh (AIUB)

[2]Department of Computer Science, Stamford University, Dhaka, Bangladesh

## ABSTRACT

A lot of sensor network applications are data-driven. We believe that query is the most preferred way to discover sensor services. Normally users are unaware of available sensors. Thus users need to pose different types of query over the sensor network to get the desired information. Even users may need to input more complicated queries with higher levels of aggregations, and requires more complex interactions with the system. As the users have no prior knowledge of the sensor data or services our aim is to develop a visual query interface where users can feed more user friendly queries and machine can understand those.

In this paper work, we have developed an Interactive visual query interface for the users. To accomplish this we have considered several use cases and we have derived graphical representation of query from their text based format for those use case scenario. We have facilitated the user by extracting class, subclass and properties from Ontology. To do so we have parsed OWL file in the user interface and based upon the parsed information users build visual query. Later on we have translated the visual query languages into SPARQL query, a machine understandable format which helps the machine to communicate with the underlying technology.

## 1. INTRODUCTION

The goal of the work presented in this paper is to develop an interactive visual query interface that works in a TOppS project scenario [1]. The task of this interface is to discover sensor information from service directory. To accomplish this, there are many problems that are needed to be overcome. E.g. What types of query interfaces are best suited for the users?

• What types of queries are posed by users when searching for sensor services?
• What is the most convenient and user friendly way for a user to query a sensor service?
• Which Machine understandable query language is to be chosen for the system?
• How does a user query get translated into a machine understandable query?

To solve the problem we need to do the following task

• The system needs to be user friendly as well as interactive so that it can be used without demanding a lot of technical knowledge from user.
• A sufficiently powerful and flexible user interface needed to be designed that is easy to learn and use.
• The user must be able to define precise and different types of queries to achieve the required service from underlying knowledge base.
• The query must be transformed into a machine understandable format. For example
SQL, SPARQL, RDQL.

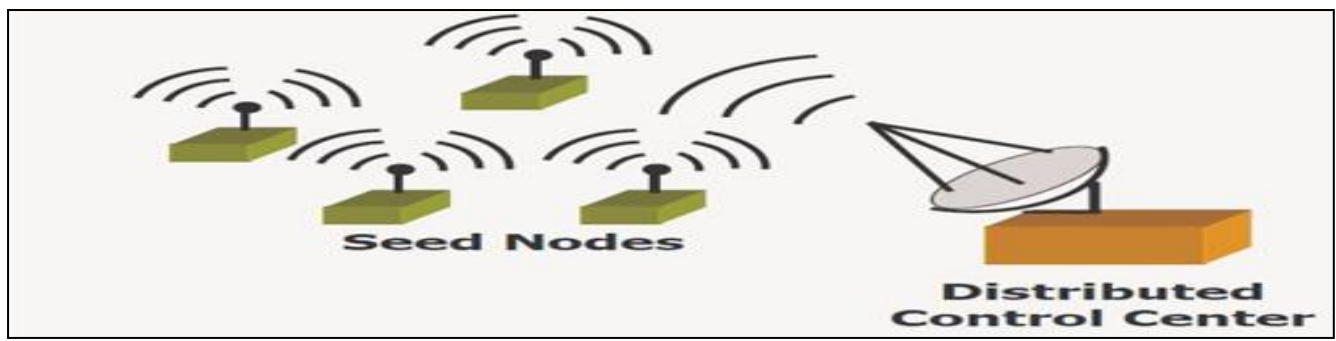

**Figure: 1.1 A seed oppnet [3]**

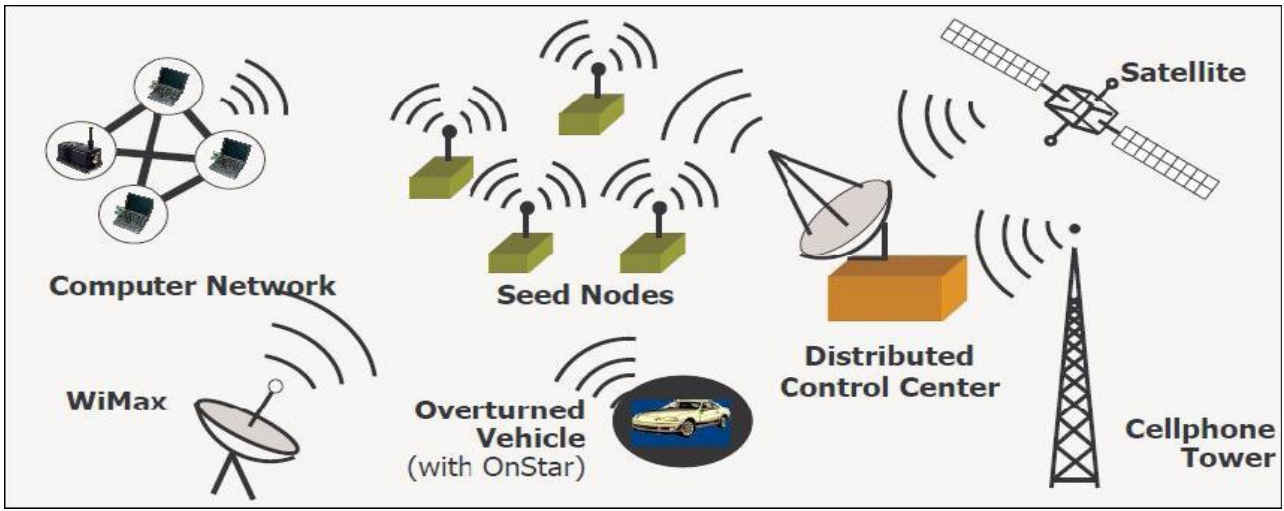

**Figure 1.2: An expanded oppnet [3]**

## 2. BACKGROUND

### 2.1 SENSOR

According to [2] a sensor is basically a device with the task to measure a physical quantity and to convert it into a signal which can be interpreted by an observer or by an instrument. For instance, a mercury thermometer converts the measured temperature into expansion and contraction of a liquid which can be read on a calibrated glass tube. A thermocouple also converts temperature to an output voltage which can be interpreted by a voltmeter.

### 2.2 OPPORTUNISTIC SENSOR NETWORK

An opportunistic network (Oppnet) [3] is a new paradigm and a new technology introduced by the four scientists of WiSe (Wireless Sensornet) Lab[3] [4] . It is a new direction within the area of computer networks. The goal of opportunistic networks or oppnets is to enable an integration of the diverse communication, computation, sensing, storage and other resources that surround us more and more.

In Figure 1.1 a seed oppnet comprise of seed nodes connecting and communicating with the distributed control center. In Figure 1.2 an expanded oppnet is shown where several nodes join the group to help the situation. The helpers that join the group are cell phone tower, WiMAX, computer networks etc.

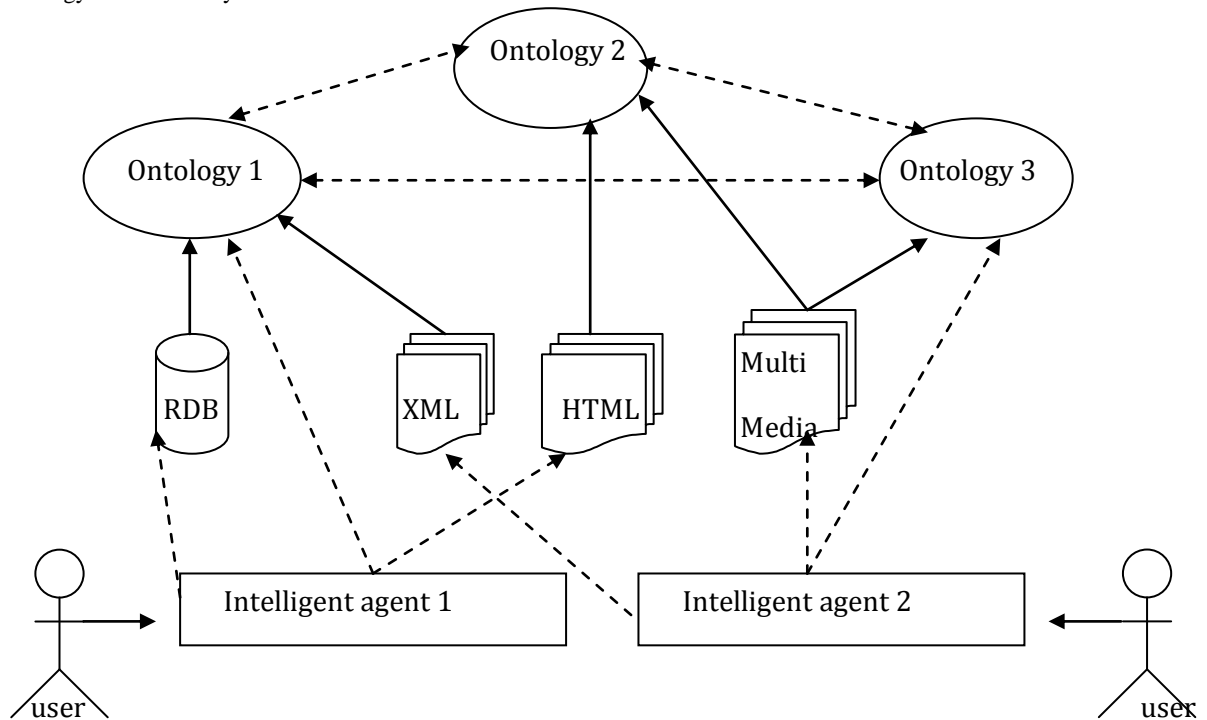

**Figure 1.3: Simple depiction of the Semantic Web [4]**

## 2.3 SEMANTIC WEB

The Semantic Web is an acquiring denotation of the World Wide Web in which the content is presented in a format that is understandable to human users as well as for software agents. This enables more efficient searching, combining and sharing of information. The Figure 1.3 [4] depicts that Intelligent agent is consuming resources like XML, HTML or multimedia resources. These resources are described in Ontologies[1] which is available in the internet and relational databases.

## 2.4 XML

The Extensible Markup Language (XML) [5] is a general-purpose specification language for documents containing structured information. As it allows the users to define their own elements so it is called as an extensible language. Its main purpose is to help information systems to share structured data, through the Internet and it is also used both for encoding documents and to serialize data [5]

## 2.5 RDF

The **Resource Description Framework (RDF)** [6] [Lassila & Swick, 1999] is a specification recommended by W3C. Its motivation is to provide a standard for meta-data model, which has come to be used as a general method of modeling information through a variety of syntax, formats.

The RDF metadata model builds on Triple pattern which is in the form of subject-predicate-object expressions. The RDF Meta data make statements about Web resources in the form of triples. The subject denotes the resource and can be represented by a Uniform Resource Identifier (URI); and the predicate denotes traits or aspects of the resource and expresses a relationship between the subject and the object. Finally, the object is a resource [7] For example, in Figure 1.4 one way to represent the notion "The camera hasResource Image" in RDF triple pattern: a subject denoting "The camera", a predicate denoting "hasResource", and an object denoting "Image".

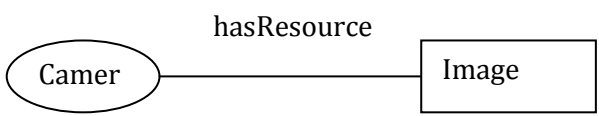

**Figure 1.4 Sample RDF representation**

## 2.6 OWL

The Web Ontology Language (OWL)[8] is endorsed by the W3C. It is a Markup language for ontologies and also sharing ontologies on the web. It is a kind of language which is used by applications to process the content of information instead of formatting or presenting information to humans. OWL appends additional vocabulary for depicting attributes and classes and also relations between classes, deeper typing of attributes, features of attributes and counted classes.

## 2.7 ONTOLOGY

Ontology is a formal representation of set of concepts in domains and also depicts the relationship between concepts. Ontologies are used in computer science to conceptualize a domain. In other words, ontologies provide a common understanding about topics that belong to a particular context.

The reasons behind the developing of ontology are as follows [9]

- To share common understanding of the structure of information among people or software agents

- To enable reuse of domain knowledge

- To make domain assumptions explicit

- To separate domain knowledge from the operational knowledge

- To analyze domain knowledge

## 2.8 SPARQL QUERY AND RDF QUERY LANGUAGE [SPARQL]

There are a number of RDF query languages, used to retrieve and manipulate data stored in a Resource Description Framework Format. Of them SPARQL is the RDF query language which is newly recommended by the W3C. SPARQL query consists of triple patterns, disjunction, conjunctions and optional patterns. Triple patterns are sequences of subject, predicate and object terms separated by white space and finished by ".".

The following simple SPARQL query taken from [10] returns all African capitals:

PREFIX abc: <http://example.com/exampleOntology#>

SELECT ?capital ?country

WHERE {

?x abc:cityname ?capital ;

　　abc:isCapitalOf ?y.

?y abc:countryname ?country ;

　　abc:isInContinent abc:Africa.

}

Variables are indicated by a '?' prefix, and bindings for?capital and ?country will be returned.

---

[1] Ontology are described in Section 2.7

### 2.9 JENA

Jena [11] is a Semantic Web framework for Java. This framework is used to build Semantic Web application. Jena offers a programmatic environment for RDF, RDFS and OWL, SPARQL and also includes a rule-based inference engine. It uses an API to evoke data from and write to RDF graphs. The graphs are represented as an abstract "model". A Model also can be queried through SPARQL query languages.

The Jena Framework consists of following elements [12]

- A RDF API
- Reading and writing RDF in RDF/XML, N3 and N-Triples
- An OWL API
- In-memory and persistent storage
- SPARQL query engine

## 3. RELATED RESEARCH

There are number of ways we can use to solve the problem. We have found that there are number of interfaces to discover sensor information. Some researchers use visual query based interface, others use natural language based interfaces and also expert users use traditional query language which to discover sensor information. For that reason we have conducted a research to compare the different options to choose the suitable one for our paper.

### 3.1 ADVANTAGES OF NATURAL LANGUAGES BASED QUERY

1. 'Text interfaces have a role to play because they are familiar to end users; benefit from very good support both on the desktop and in web interfaces, and are easily available on all types of devices' [13] .

2. 'The evaluation conducted [14] contains a usability test with 48 end-users of Semantic Web technologies, including four types of query language interfaces. They concluded that NL interfaces were the most acceptable, being significantly preferred to menu-guided, and graphical query language interfaces. Despite being preferred by users, Natural Language Interface (NLI) system are not very frequent due to the high costs associated with their development and customization to new domains, which involves both domain experts and language engineers'[13] .

3. 'Probably due to the extraordinary popularity of search engines such as Google, people have come to prefer search interfaces which offer a single text input field where they describe their information need and the system does the required work to find relevant results' [13] .

A few examples of Natural language systems are SemSearch, Aqualog, Orakel, ONLI, Querix.

### 3.2 DISADVANTAGES OF NATURAL LANGUAGES BASED QUERY

While the querying paradigm based on natural language is generally deemed to be the most intuitive from a usage point of view, it has also been shown to be the most difficult to realize effectively. The main reasons for this difficulty are that: [15]

1. Natural language understanding is indeed a very difficult task due to ambiguities arising at all levels of analysis: morphological, lexical, syntactic, semantic, and pragmatic (compare [Androutsopoulos et al., 1995, Copestake and Jones, 1989].

2. A reasonably large grammar is required for the system to have an acceptable coverage.

3. The natural language interface needs to be accurate.

4. The system should be adaptable to various domains without a significant effort.

### 3.3 USING FORMAL LANGUAGES AS A INPUT QUERY

1. Protégé [16] provides the Query Interface, where one can specify the query by selecting some options from a given list of concepts and relations. Alternatively, advance users can type a query using a formal language such as SPARQL. Facilities of this type are that it gives the maximum level of control but are most appropriate for experienced users.

2. 'While typing queries in formal languages provides the greatest level of expressivity and control for the user, it is also the least user-friendly access interface' [13] .

3.'Query languages have complex syntax; require a good understanding of the representation schema, including knowledge of details like namespaces, class and property names. All these contribute to making formal query languages difficult to use and error prone' [13] .

4. 'The obvious solution to these problems is to create some additional abstraction level that provides a user friendly way of generating formal queries, in a manner similar to the many applications that provide access to data stored in standard relational databases'[13] .

### 3.4 VISUAL QUERY LANGUAGES

People can use natural languages to interact with the system. But it is seen that natural language bases query is too much complex and it is hard for the system to process natural based query. So we found two solutions for these, either we can use the simplified natural language or we can use the visual query languages.

Visual Query languages [17] are not limited to characters and symbols, but includes all possible interactions between a human and a machine, i.e. mouse movements, buttons, graphics, etc. In general, users easily accept graphical interfaces that follow the normal windows standard of interaction with mouse, keyboard and graphical display.

The input part of the user interface is how the user specifies a query. It has to be flexible so that the user can formulate precise queries, yet easy and efficient to use. In the command and control situation the aspects that are required to accomplish this [17] .

In [17] researchers use the visual query languages for user interaction since this type of approach is very familiar to end user and they also realize that it will be suitable for command and control environment. The system is designed in a way that the user can give their request with terms that are very natural and simple for them. The interface is so user friendly that the users who have no previous knowledge about the system can easily interact with the system.

## 3.5 NITILIGHT

As we are building an interface for semantic queries we have also looked at existing tools which help users to build graphical semantic query. NITELIGHT [18] is one of them. It is developed by Intelligence, Agents, Multimedia research group of University of Southampton, England.

NITELIGHT[18] is a graphical tool is based on the SPARQL query language specification. The tool supports end users by providing a set of graphical notations that represent semantic query language constructs. This language provides a visual query language counterpart to SPARQL which is called vSPARQL. It also supports an interactive graphical editing environment which includes ontology navigation capabilities with graphical query visualization techniques.

The idea behind the development of the tool is that it supports the user in building syntactically valid queries. It serve to constrain or guide editing actions so as to mitigate against the risk of lexical or syntactic errors.

The tool enables users to create SPARQL queries using a set of graphical notations and GUI-based editing actions. The tool is intended primarily for users that already have some familiarity with SPARQL; the close correspondence between the graphical notations and query language constructs makes the tool largely unsuitable for users who have no previous experience with SPARQL.

## 3.6 iSPARQL

The iSPARQL[19] is open source software which is known as OpenLink iSPARQL.The iSPARQL Visual Query Builder supports the user with respect to the specification of all SPARQL query result forms (i.e. SELECT, CONSTRUCT, etc.). It also supports the creation of optional graph patterns as well as UNION combinations of graph patterns. This tool is very rich to create graphical semantic query and SPARQL translation.

This tool also provides VQL for their graphical representation. The user can process ontology from Tree view manner and build the query bases on class and concepts of the Ontology. One of the advantages of iSPARQL is that it can built very complex semantic graphical notation.

## 3.7 SPARQL

The reason that mainly inspire us to choose SPARQL query language are [20] -

- There are many distinct database technologies in use, and it's of course impossible to dictate a single database technology at the scale of the Web. RDF (the Semantic Web data model), though, serves as a standard lingua franca (least common denominator) in which data from disparate database systems can be represented. SPARQL, then, is the query language for that data. As such, SPARQL hides the details of a saver's particular data management and structure details. This reduces costs and increases robustness of software that issues queries.

- SPARQL saves development time and cost by allowing client applications to work with only the data they're interested in.

- SPARQL builds on other standards including RDF, XML, HTTP, and WSDL. This allows reuse of existing software tooling and promotes good interoperability with other software systems.

- SPARQL results are expressed in XML: XSLT can be used to generate friendly query result displays for the Web.

- It's easy to issue SPARQL queries, given the abundance of HTTP library support in Perl, Python, Php, Ruby, etc.

## 4.DESIGN

We have decided to use the Visual query language because we believe that it is more user friendly. One of our problems to investigate was '*What types of queries are posed by user for searching the sensor services?*' Therefore we choose a visual query to solve the problem. Another of our problems stated in Chapter 1 is '*Which Machine understandable query is to be chosen for the system*'. After the related research in previous Chapter we have decided to use SPARQL as our machine understandable query language.

### 4.1 GRAPHICAL NOTATIONS

SPARQL queries exploit the triple-based structure of RDF models. A graph-based query representation consists of a sequence of nodes and links. It can be used to represent the core of most SPARQL queries, for example the basic triple patterns that are matched against the RDF data model. The nodes in this case correspond to the subject and object elements of an RDF triple; the links correspond to RDF predicates.

In our paper we use a circle to represent the subject and a rectangle to represent the object. Predicates are shown as simple lines and associated with a label that indicates the predicate name or the name of a query variable. Directional arrows point to which node represents the subject and which node represents the object in a triple pattern.

## 4.2 GRAPHICAL PRESENTATION

In this section we have showed three main elements of our designs. The first one is user expression in natural languages, visual representation of natural languages and corresponding SPARQL queries according to the visual language. The Visual representation we have used in our paper is depicted in Figure 3.1 and Figure 3.2.

For example a user wants to retrieve Information from TOppS framework. The user natural language query is as follows.

Text representation- "*I want the Image and Video from Camera Sensor.*"

Here a user wants Image and Video from Particular Camera sensor. The visual representation of this query is described in Figure 3.1.

Where **?²x=Camera Sensor.** As user is searching for Image and Video, **Image** and **video** class is needed to design visual representation. User also has to insert corresponding properties. In this scenario Camera class has properties **hasResourceType**.

The SPARQL follow the triple based structure of RDF. Here **?x** is the subject, **tp:hasResourceType** is predicate and **?Image,?Video** is object. The corresponding SPARQL of the Figure 3.1 is

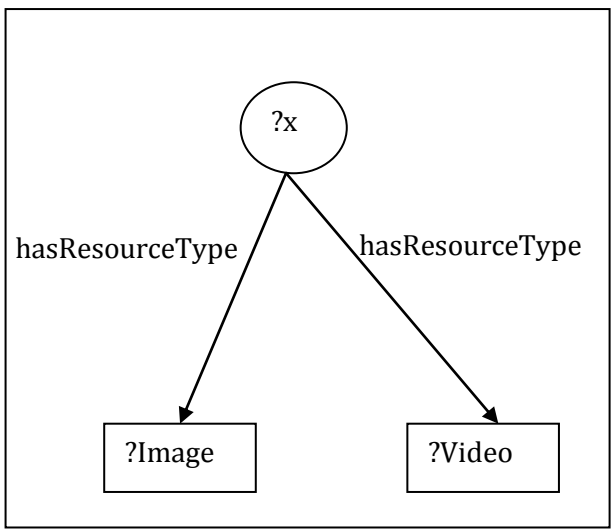

**Fig 3:1 Graph based representation of semantic query**

**Select ?Image,?Video**

**Where**

**(**

---

**?x tp:hasResourceType ?Image;**

**?x tp:hasResourceType ?Video;**

**)**

Let's consider a complex scenario for the case study. For example in an Alarm Scenario, an alarm is triggered when someone enters into a room. The camera captures the image of the person when the motion detector triggers an alarm event. In this scenario two tasks happened; one is someone enter the room and the alarm event is triggered by the motion detector. The next task is for the camera to turn on and captures the image. The natural language query for this scenario is

Text Representation- **I want image when someone enter the room**

The graph representation of an alarm scenario is shown in Figure 3.2. Where **?x** denotes camera sensor. It has properties **has_uri, has_resource and has_location.** This scenario also contain **class ?y** which denotes the URL location of the image. **Image** and **room** are separate class. **?z** denotes motion detector class which has properties get_detection. **Binary** is another class which represents the value in 1 or 0.

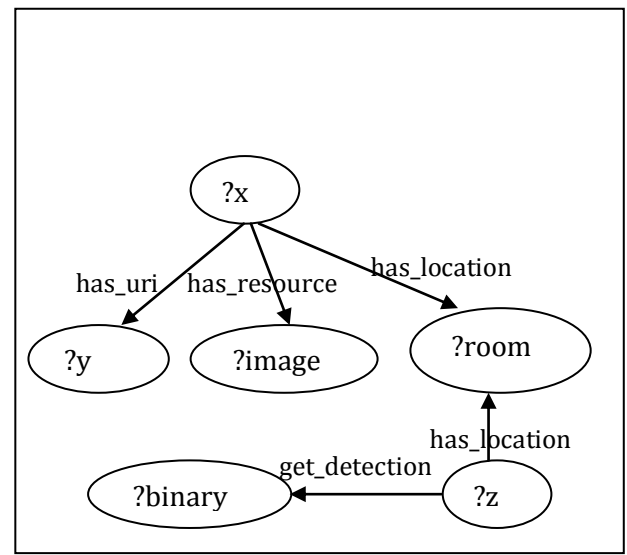

**Fig 3:2 Graph semantic query for alarm scenario**

The SPARQL query will be like this as user is searching for image. Basically he needs the URL of the Image.

**Select ?y**

**Where**

**(**

**?x has_uri ?y;**

**?x has_resource ?image;**

**?x has_location ? room ;**

**?z has_location  ?room ;**

**?z get_detection 'true';**

**)**

## 4.3 USER WORK FLOW

The Steps the users have to follow in our paper are below

At first the user goes to the file menu and chooses the OWL file from the directory.

The parser parses the OWL file and extracts the classes, subclasses and properties.

The extracted classes and properties are shown in the interface.

Based on the classes and properties shown in the interface the user will build the semantic graphical query with the options provided in the design canvas.

Finally the graphical representation will be translated to SPARQL and displayed to the user.

The user can save the query and also send the query to the rest of the framework to retrieve information.

## 4.4 DESIGN QUERY EDITOR PROTOTYPE

To test and evaluate the features of graphical query we have designed an editor where user can give the query .

The editor consists of

Ontology parser(OWL Parser)

Query design canvas

SPARQL query viewer

### 3.4.1 Ontology parser (OWL Parser)

To facilitate the process of query formulation and query specifications our editor consists of an ontology parser. The users load the ontology by browsing the file from any locations of the computer. The loaded ontology results in the enumeration of classes and sub classes. The ontology parser also facilitate with the enumerations of properties. The user can watch the classes and properties from parser canvas and can build the query according to his wish.

### 3.4.2 Query design canvas

This is the core design canvas for user interactions and query constructions. This section consists of two parts, one is query canvas part and another one is query tool part.In the query canvas part the user gives the user questions. The graphical SPARQL query is portrayed in this canvas. The Query builder part consists of several options. In this part the user give the information of the classes and properties that they want to build the graphical query. During the period of giving information validation checking is done to verify that the information is present in the currently loaded ontology or not. After performing validation checking the user can create relations and create query by using the corresponding button. There is also a refresh button, the user can use it to clear the canvas and redraw the query again.

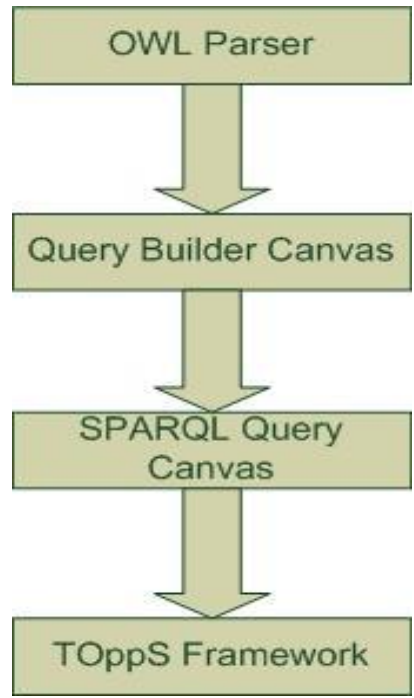

**Figure 3.3 Graphical query Builder Framework**

### 3.4.3 SPARQL query canvas

The SPARQL query viewer displays the SPARQL query in this canvas and this query is updated dynamically according to the graphical query. Right now there is no option to edit this query but it is updated according to the graphical query built by the user. If the user wants to change any syntax the user has to change the graphical query and its changes are reflected in the SPARQL query viewer.

To communicate with the TOppS system the SPARQL query is sent via a web service interface to retrieve information from the registry

## 5.IMPLEMENTATION

The implementation of the project was carried out at FOI, Totalförsvarets forskningsinstitut, as part of a research project on developing dynamic services for opportunistic sensor networks. As a part of the research scheme we have developed Graphical user interface part.

In this chapter we will briefly explain the implementation issues related to our paper.

### 5.1 PLATFORM

Following is the development platform for the Query builder

Operating System: Windows

Java: JDK 1.5

Net Beans IDE 6.5

Jena 2.5.7

### 5.2 QUERY BUILDER

Following in Figure 4.1 is the interface of Graphical user interface (GUI).

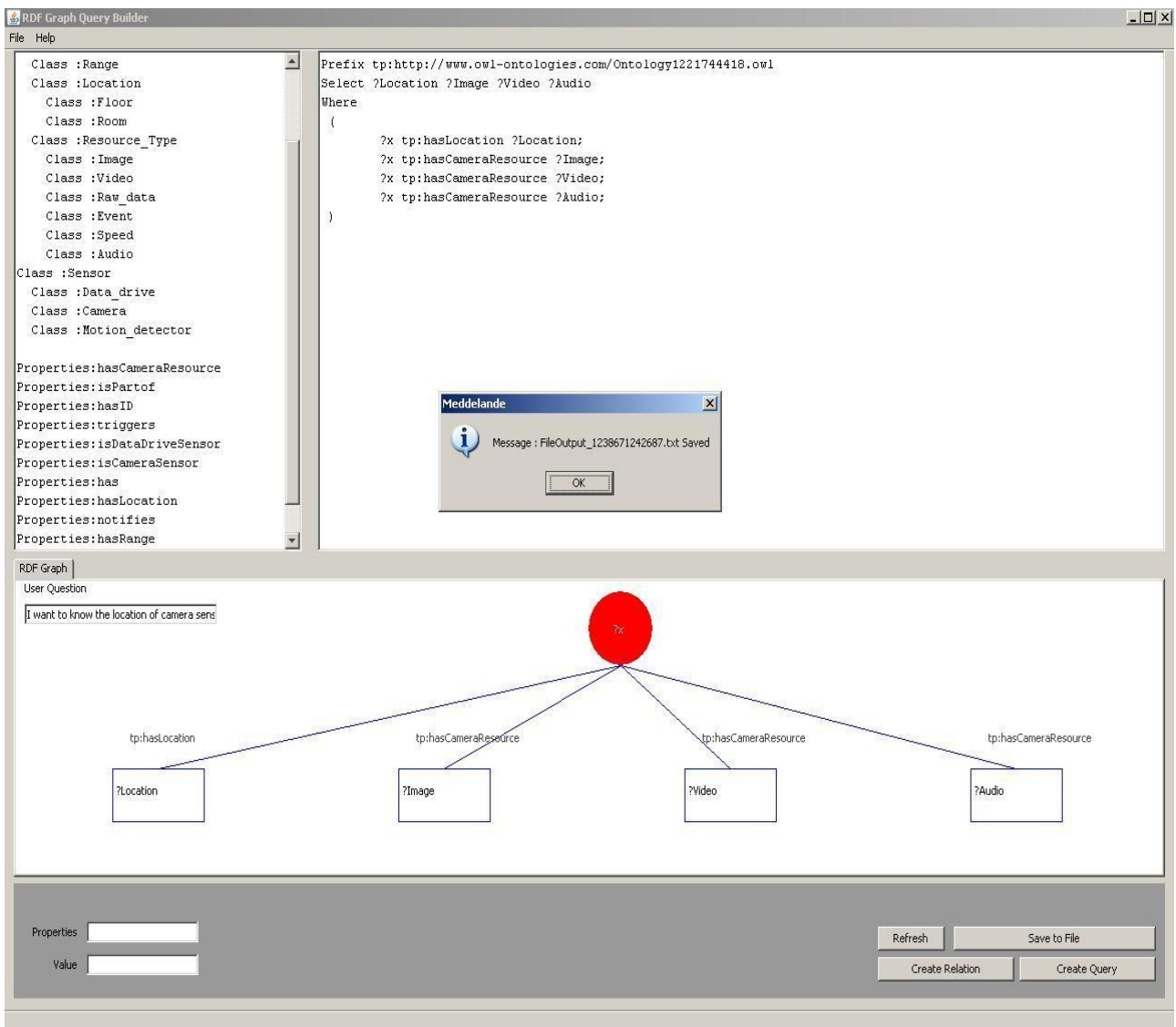

**Figure 4.1 Query Builder Interface with a query**

The complete list of classes, sub classes and properties of any OWL is displayed on the upper left part of the interface. The upper right part shows the translated SPARQL query. In the middle part the graphical query is build. At this stage Simple graphical queries are built which is explained detail in testing and evaluation Chapter. The user can add up to twelve graphical nodes in this interface. The bottom part consists of user input tools. In this part the user can specify the properties and classes which they want to use for the particular ontology. Here the user also gets the chance to clear the canvas and draw a graphical query again and finally the user can save the SPARQL query with the natural language user question in a txt file.

## 5.3 SENSOR ONTOLOGY
For our system requirement we have also designed and developed Testing Sensor ontology for our work. The developed Sensor Ontology with classes and sub classes is depicted in Figure 4.2.

## 6.TESTING AND EVALUATION
Testing is the most important issue in the software development life cycle. To check the proper functionality of the software we have tested our implemented prototype with different input and verified the result of the output. The detailed description of the testing is described in this portion. In the design Chapter we have designed the Graphical Query Builder. In this testing phase we have checked the functionality, verified the input and made plans for the usability testing of implementation.

### 6.1 ONTOLOGY READING TEST
In the design Chapter we have described the Ontology Parser (OWL Parser). It is one of the three components of our systems. To test the functionality of the OWL parser we have tested it with different OWL files. The samples we have used to test OWL parser are Pizza Ontology[21] ,Wine Ontology[22] and our implemented Sensor Ontology. The results that we have got from the Sample OWL file are depicted in Figure 5.1. As we can see the interface is able to successfully read and parse the different ontology OWL files and output the classes and properties.

### 6.2 GRAPHICAL QUERY TESTING
The second component of the system is the graphical query canvas. To test this component we have employed some use cases described in the followings experiments.

The classes in the ontology are designed such that there is a distinction between the sensors devices and the data that the devices have on them. The sensor class describes the different types of Sensors in the system. The data class describes different data categories that the sensor has. In the figure 4.2 we have showed the relations between classes with properties. For example from the figure we can see that the camera sensor class has properties 'hasCameraResource' Image, **Audio** and **Video.**

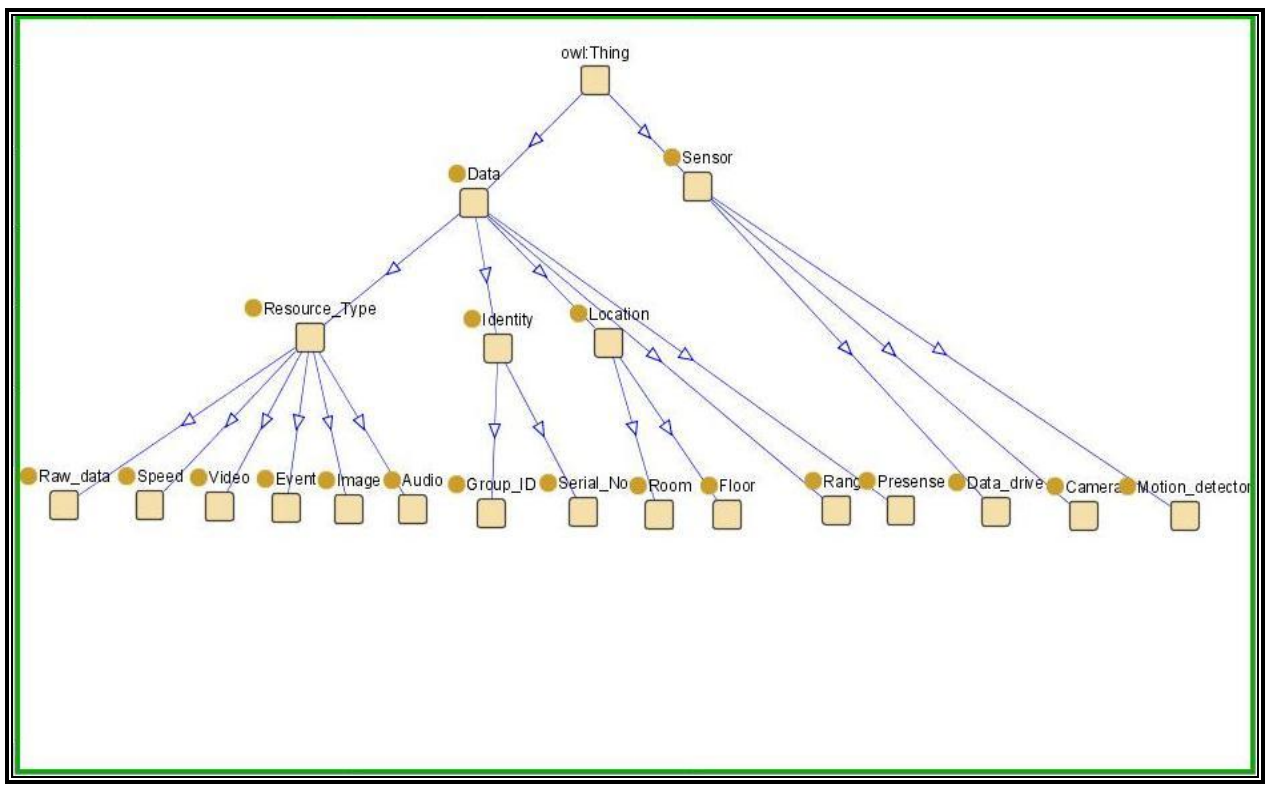

**Figure 4. 2 Sensor Ontology With classes and Sub classes**

## Experiment 1

The natural language query that may be asked by the user is as follows *"Find the Image from the Camera Sensor"*.

The corresponding Graphical query is shown in Figure 5.2.

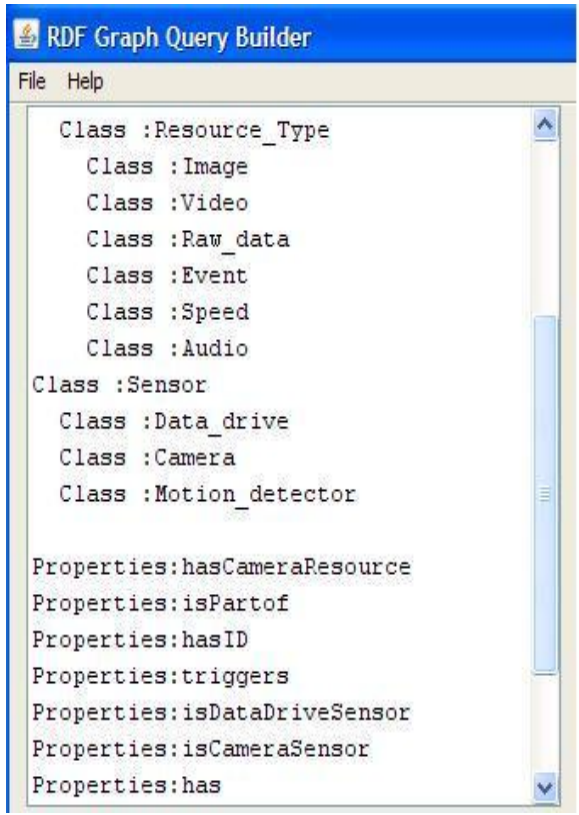

**Figure 5.1 Test Result with Sensor Ontology**

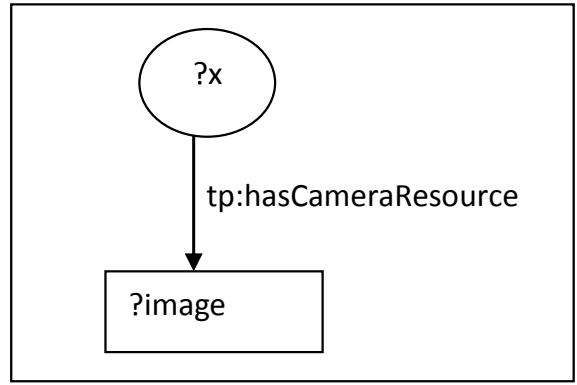

**Figure 5.2 Graphical Query for Sample data 1**

Here ?x denotes the Camera Sensor, ?image is the image (class value) that the user asks for from the Camera Sensor. hasCameraResource is the property value that is needed to build the graphical query and tp is the namespace

The corresponding SPARQL query is as follows

**Select ?image**

**Where (**

**?x tp: hasCameraResource ?image ;**

**)**

The output that our program gives for Experiment is given in figure 5.3

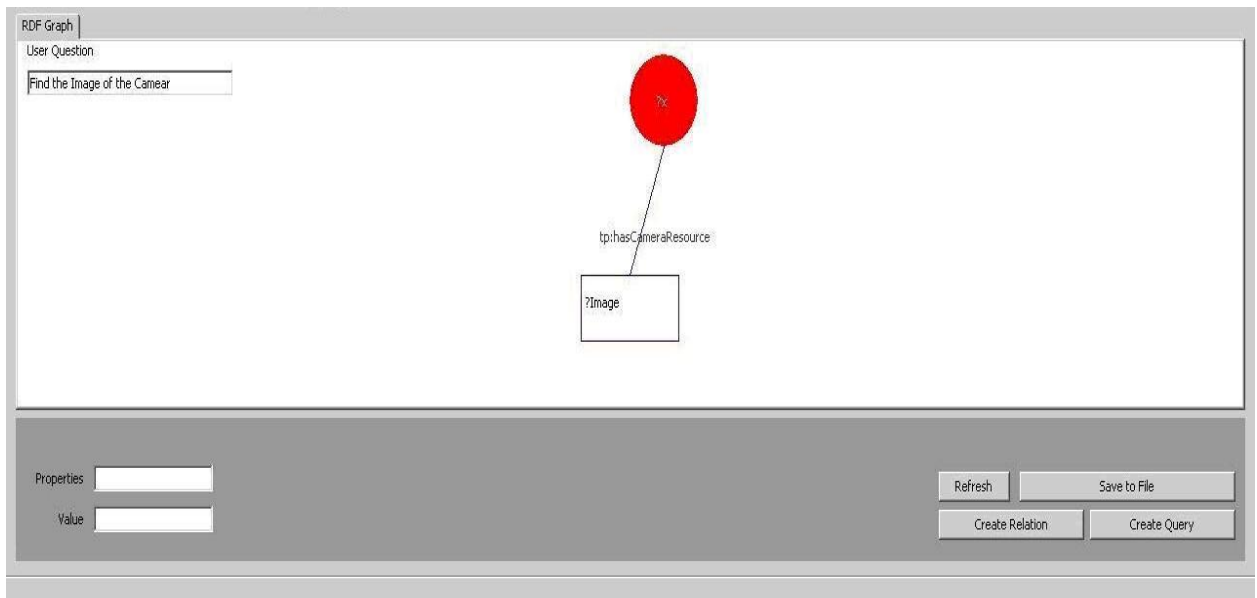

**Figure 5.3 Graphical Query output for Sample data 1**

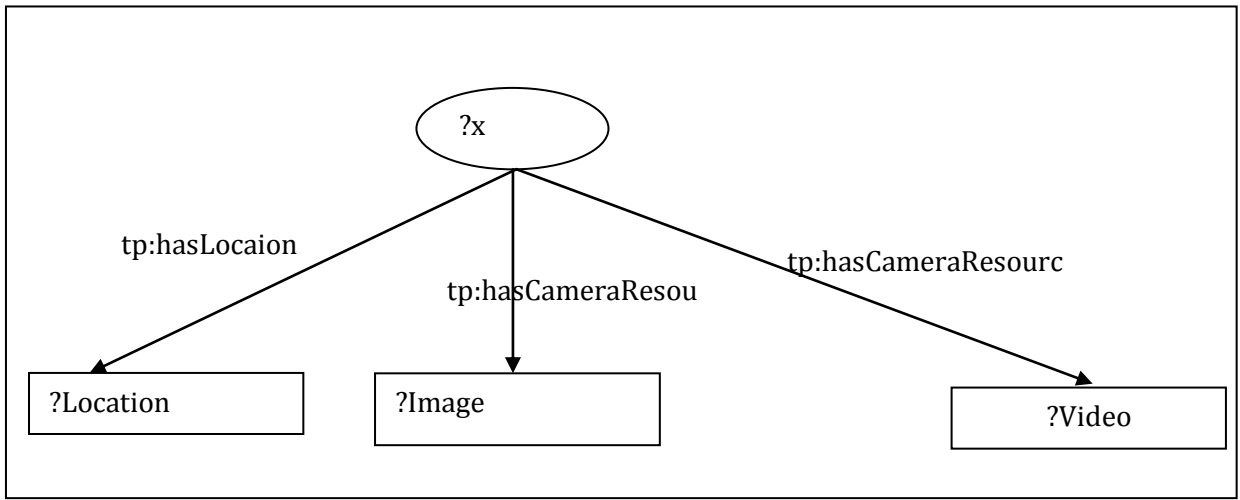

**Figure 5.4 Graphical Query for Sample data 2**

**Experiment 2**

A more complex question that a user may ask is as follows

*"I want to know the location of camera sensor and image, video from camera sensor".*

The corresponding graphical query that our system draws is painted in Figure 5.4

Here **?x** denotes the Camera Sensor. The user is asking for Location of the camera and also the Image and Video from camera so user has to use the Class of Location, Image and Video which is represented as **?Location, ?Image,? Video**. The properties value is represented as **hasLocation, hasCameraResource**.

The corresponding SPARQL queries is as follows

**Select?Location ?image ?Video**

**Where**

**(**

**?x tp:hasLocation ?Location;**

**?x tp:hasResourceType ?Image;**

**?x tp:hasResourceType ?Video;**

**)**

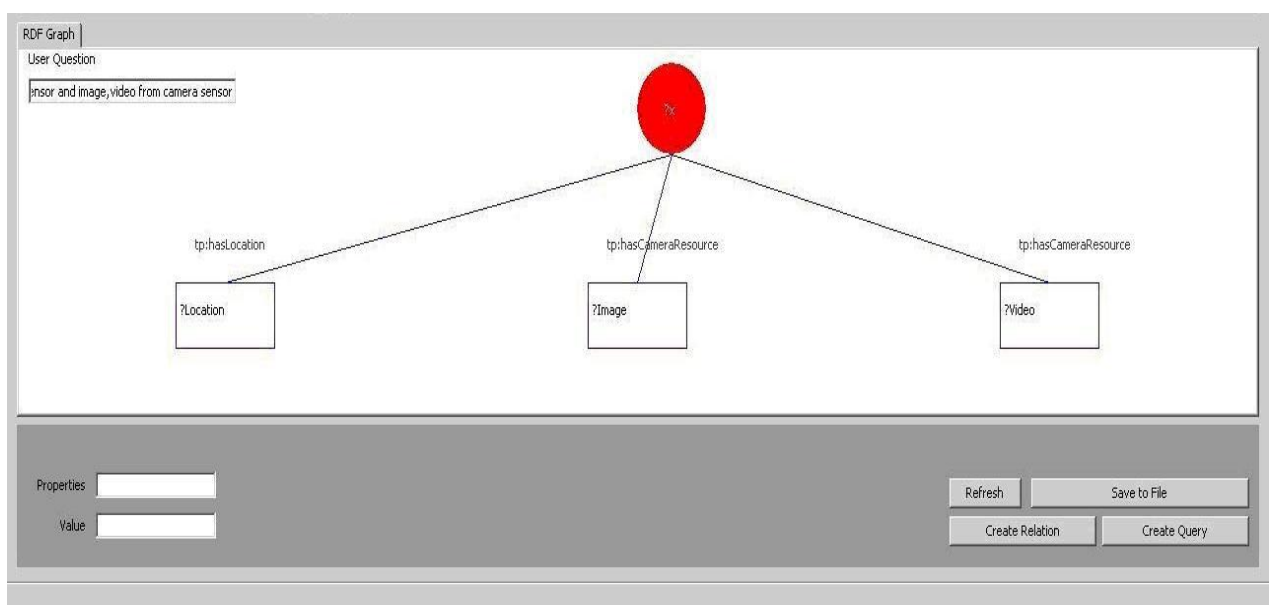

**Figure 5.5 Graphical Query output for Sample data 2**

**Experiment 3**

The user may also give some advanced level query where the user may not use keywords to express his query. For example

*"I want image from room when someone enter the room"*

The corresponding graphical query is shown in Figure 3.2

Here **?x** denotes camera sensor as it will return image. **?y** is the URL of the Image. **?z** is the motion detector sensor. **?Image, ?Location,?binary** is the class value. **has_uri, has_location, has_resource, get_detection** are the properties value.

**Evaluation of Experiment**

The output that we get in Figure 5.3 and Figure 5.5 is semantically correct as it follows the ontology structure. The user can build this structure using the Sensor Ontology . The Sensor Ontology which is described in Chapter 4 shows that **Camera** class has properties **hasResourceType** and it is associated with the resource type class **Image**, **Video** and **Audio**. This class has another properties **hasLocation** which is associated with **Location** class**.** The detailed view of the Sensor Ontology is depicted in Figure 4.4. In our experiments test 1, test 2 and test 3 we have found that our implemented prototype is following the structure of the Sensor Ontology . Finally we can say that our implemented prototype is semantically correct.

## 6.3 SPARQL QUERY TESTING

The translated SPARQL queries that we got from experiments, see the Figure 5.2 and Figure 5.4 follow the triple based structure of SPARQL queries. We compare the SPARQL queries with the Standard [23]  to test the syntax. The sample SPARQL query that we get from the definition [23]  is as follows :

PREFIX foaf:  <http://xmlns.com/foaf/0.1/>

SELECT ?name ?mbox

WHERE

  {

    ?x  foaf:name  ?name .

    ?x  foaf:mbox  ?mbox .

  }

 The Translated SPARQL query of our system is shown in Figure 5.6.

We have compared the result of Figure 5.10 with the resource [23]  sample query, and find that our system is providing SPARQL queries with the correct syntax.

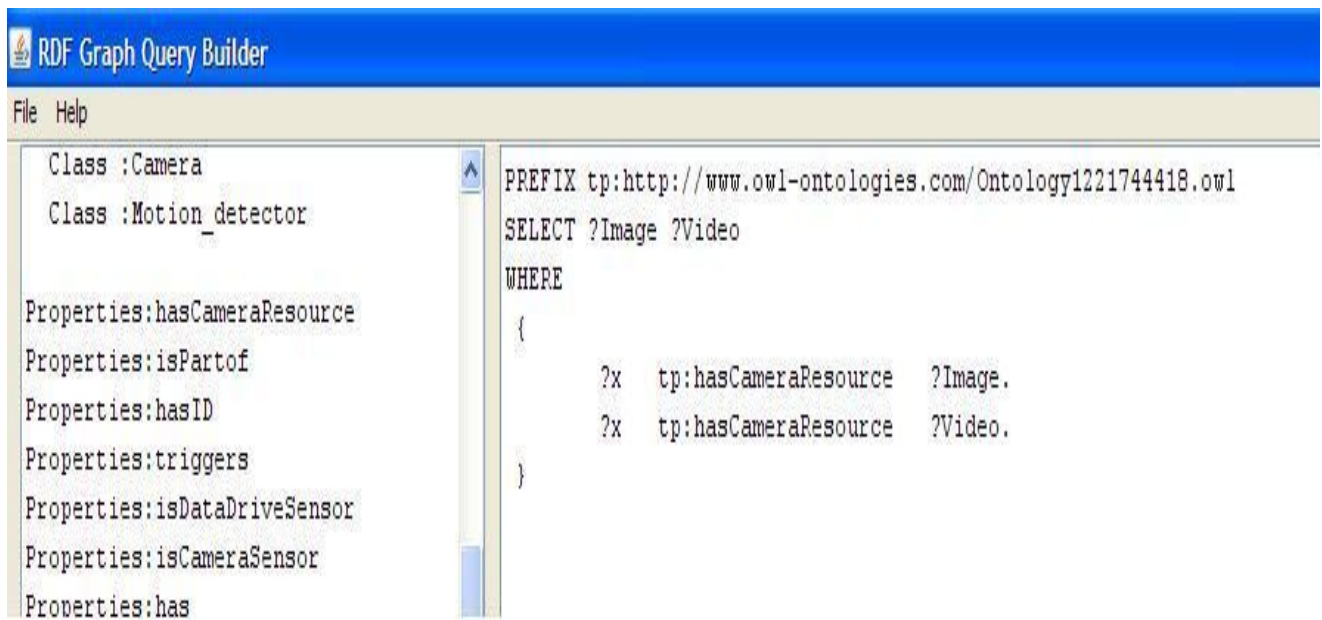

**Figure 5.6 Translated SPARQL query**

## 6.4 USABILITY TESTING

At this stage we have not performed any usability testing. However our aim is to undertake such testing in the near future. Specific focus for testing includes the general usability of the tool, the ability of the tool to support users with regard to query sending to the underlying technology. Our focus also compare our prototype with NITELIGHT [18] and iSPARQL[19] We also have planned to use a questionnaire with the following questions:

- How easy did you think it was to use the graphical query interface on a scale of one to five where one is difficult and five is easy?

- How user friendly do you think the interface appearance is on a scale of one to five where one is not user friendly and five is very user friendly.

- How easy did you think it was to load the Ontology file on a scale of one to five where one is difficult and five is easy?

- How easy did you think to build a graphical query in the design canvas on a scale of one to five where one is difficult and five is easy?

- How easy did you think it was to understand the relation between the graphical query and SPARQL query on a scale of one to five where one is difficult and five is easy?

## 6.5 COMPARISON WITH NITELIGHT

NITELIGHT is a Semantic query editing tool. It supports visual query building and different options of SPARQL query. Our work is similar to NITELIGHT but we have a different scope than NITELIGHT. Our prototype is dedicated to Opportunistic sensor network. NITELIGHT have their on vSPARQL which doesn't match in Opportunistic sensor network. More over NITELIGHT is not an open source tool that can be used. The NITELIGHT tool is an editing tool for SPARQL query, where as our prototype fully designed for communicating with the TOppS framework. NITELIGHT tool does have any facility to express user questions in natural language where as in our prototype user is expressing his natural language query in the interface and then according to the natural language statement the user get the class and properties value in the OWL parser and can build the graphical query. NITELIGHT tool is designed for expert users who are knowledgeable in the field of semantic web. We have tried to design the system more user friendly way that a normal user who has some knowledge about query formulation can use our system. In our prototype we have also proposed some advanced case study for example **Alarm Scenario** that can be solved by our prototype. The most important issue with our prototype is that our prototype is not an editing tool; it builds the visual query, translating to SPARQL and finally sends the query to the TOppS framework. Our aim is that the users who have no previous knowledge in SPARQL query can use our system to send the query to the web service accepting SPARQL queries. The user doesn't have to write the code manually where as the system translate the query for the user.

iSPARQL is another graphical Semantic Query editing tool. The difference between our work and iSPARQL is that it doesn't have support for communication with other framework. It also differs in the aspect of loading the Ontology file. In our prototype we have used the menu based structure to load the ontology file where as iSPARQL use the tree view structure to load the ontology file[18] . In our system user is giving natural language query and then from that he can build the graphical query where as in iSPARQL doesn't have natural language facility.

## 6.6 Evaluation

We have designed graphical query based on the triple based structure of RDF. Manually we have also designed some experimental graphical query and translation version of SPARQL query. In the testing period our main goal was to monitor the following options in our system.

- Processing of OWL file.
- Representation of Class and Properties from OWL file.
- Building graphical query according to the user input.
- Building the corresponding SPARQL query from the graphical query.

From the testing result we have found that our implemented system can process any existing OWL file and it can extract class and properties from OWL file and display the classes and properties to the user. In the testing we have also established that our implemented system is working with the testing experimental data. It can build the graphical query according to the design and it can translate the graphical query to SPARQL query. The specification we have designed in our designing chapter is working fine with our implemented prototype. From the comparison of NITELIGHT and iSPARQL we see that both this are editing tool where as our prototype designed to communicate with framework.

Right now we have no option in our implemented prototype to use multiple co-depended triples. It can process only one level of query. So it cannot process queries like the experiment 4 data. In the future there is a scope and option for expanding the implemented prototype.

## 7.SUMMARY AND FUTURE RESEARCH

The research problem was to build graphical query and to dynamically translate it into SPARQL query. This was researched in detail and implemented as a case study in the Java environment. The implementation details are as follows:

- A parser was written in Java using the Jena API to process the ontology file (OWL).It fetch information from the OWL file and presents it to the user.

- A graphical canvas was provided to the user to build the graphical RDF query. The user looks at the ontology information and builds the graph. A validation checking was done in this canvas, so that the user can put the authentic ontology information.

- A SPARQL query canvas was provided which dynamically translate graphical query to SPARQL query. Then this information is displayed in the user canvas.

- The user can save the user questions with SPARQL query in the text file.

## 8.CONCLUSION

In this paper we have successfully designed the Interface for the TOppS graphical query builder. The GUI is developed having the functionality of user friendliness, ease of use, functionality to build more precise queries and translating it to SPARQL. The problem that we have stated in Chapter 1 was to build a user friendly interface where user can provide different kinds of query and as a requirement the query should be translated in a format that can be understood by the underlying technology. At the end of the paper we have achieved the following goals.

- Investigated query types and query languages and choose the most suitable for the task.
- Graphical User Interface (GUI) for query building has been designed and implemented.
- A canvas is implemented where user can give their queries. To achieve this we process a OWL file using Jena. So that the user can build graphical query according to the information presented in the OWL file.
- As a requirement, we have translated the graphical query to SPARQL query so that the underlying technology can understand the user query.

We believe that the goal which we proposed is achieved at the end of this work.

## 9.FUTURE WORK

The implementation is not yet fully completed and following are some of the ways to extend it.

- **Communication –** As a requirement of the paper we have translated the graphical query to SPARQL query. Right now we are saving SPARQL query and natural language user question in a text file. In future a web service interface can be introduced for communication with the TOppS framework.

- **Complex query**-In the paper we have designed different types of query. At present we have developed one level of query due to the time limitation. In the future this should be extended to handle more complex queries.

- **User Interface**- In the paper we have developed the graphical interface for user. In future it could be made more user friendly. For example, instead of giving input by keyboard the user can drag and drop class and properties from the canvas to build the graphical query.

## 10.REFERENCES

[1] Tim Fristedt, Marialena Garcia Lozano,Pontus Horling,Farshad Moradi,Johan Pelo,Peter Sigray 'Service oriented Opportunistic Sensor networks(TOppS Project)- Feasibility study year 2007'.

[2] Wikipedia The Free Encyclopedia. Sensor. http://en.wikipedia.org, [accessed Apr 2009].

[3] Lilien, L. Gupta, A. Zijiang Yang. (2007). Opportunistic Networks for Emergency Applications and Their Standard Implementation Framework.

[4] Wikipedia the Free Encyclopedia. Semantic Web. http://en.wikipedia.org, [accessed Feb 2011].

[5] Wikipedia the Free Encyclopedia. XML. http://en.wikipedia.org, [accessed Feb 2011].

[6] Decker, S. et al., (2000) '*The Semantic Web: The Roles of XML and RDF*' Internet Computing IEEE (Vol. 4, No. 5) pp. 63-74.

[7] Wikipedia the Free Encyclopedia. RDF. http://en.wikipedia.org, [accessed Feb 2011].

[8] Wikipedia the Free Encyclopedia. OWL. http://en.wikipedia.org, [accessed Feb 2011].

[9] Natalya F. N. & Deborah L. M.(2001) `Ontology Development 101: A Guide to Creating Your First Ontology'. Stanford Knowledge Systems Laboratory Technical Report KSL-01-05 and Stanford Medical Informatics Technical Report SMI-2001-0880.

[10] Wikipedia the Free Encyclopedia. SPARQL. http://en.wikipedia.org, [accessed Feb 2011].

[11] Wikipedia the Free Encyclopedia. Jena. http://en.wikipedia.org, [accessed Feb 2011].

[12] A Semantic Web Framework for Java - 'http://jena.sourceforge.net/' Last visited September 3, 2008.

[13] Bontcheva, K., Damljanovic, D.& Tablan, V (2008)' *A Natural Language Query Interface to Structured Information* ' In: Proceedings of the Fifth European Semantic Web Conference (ESWC 2008), Spain.

[14] Kaufmann, E., & Bernstein, A. (2007): '*How useful are natural language interfaces to the semantic web for casual end-users?*' In: Proceedings of the Forth European Semantic Web Conference (ESWC 2007), Innsbruck, Austria.

[15] Cimiano, P., et al., (2007) '*ORAKEL: A Portable Natural Language Interface to Knowledge Bases*' Technical Report, Institute AIFB, University of Karlsruhe.

[16] Noy, N., et al., (2001) '*Creating Semantic Web Contents with Prot_eg_e-2000*' IEEE Intelligent Systems 16(2) 60-71.

[17] Karin Camara 'A visual query language served by a multi-sensor environment'Master thesis submitted at Linkoping University.

[18] Russel, A. et all, (2008) '*NITELIGHT: A Graphical Tool for Semantic Query Construction*'. In: Semantic Web User Interaction Workshop (SWUI 2008), 5th April 2008, Florence, Italy.

[19] OpenLink iSPARQL-'http://demo.openlinksw.com/isparql/' [accessed Feb 2011].

[20] Technical Speaking Blog-'http://thefigtrees.net/lee/blog/2008/01/why_sparql.html' [accessed Feb 2011].

[21] Pizza Ontology v 1.5 'http://www.co-ode.org/ontologies/pizza/2007/02/12/'[accessed Feb 2011].

[22] Wine Ontology 'http://www.w3.org/TR/owl-guide/wine.owl'[accessed Feb 2011].

[23] W3C Recommendation SPARQL Query Language for RDF 'http://www.w3.org/TR/rdf-sparql-query/' [accessed Feb 2011].